# Life on a tidally-locked planet

## Introduction

A tidally-locked planet in its orbit around a star keeps the same face towards the star. This happens when the rotation period of the planet around its own axis becomes equal to its revolution period around the star. The rotation and revolution periods, even if initially different, could get synchronized over time due to tides on the planet because of the gravitational effect of the star, hence the term tidal-locking. Of course instead of a planet and star it could be any other gravitationally bound pair of astronomical bodies, like Earth and Moon.

Moon keeps the same face turned toward Earth at all times. This is the effect of the tides that might have been caused on Moon due to the gravity of Earth. Over time Moon's rotation period has become synchronized with its period of revolution around the Earth. Tides caused by both Moon and Sun are slowing down Earth's rotation too and the day is lengthening day-by-day. In a very far-off distant future the day would become much longer, first a month-long, when Earth gets tidally locked to Moon and then perhaps a year-long, when Earth becomes a tidally-locked planet with its rotation period equal to its revolution period around Sun. Not infrequently, a *leap* second has to be added in the standard time keeping, based on atomic clocks, to keep in synchronism with this slow lengthening of the day.

Tidal locking will do a number of things to Earth's ecosystem as the front side starts getting all the sunlight, and the back side slowly freezes. In fact an extremely important question that comes to mind in this connection is - could life be sustained on such a planet which has one side always facing the Sun (temperatures much above $100^\circ$ C, the boiling point of water) and the other darker side (howling winds at below $-100^\circ$ C)? We do not discuss here the miniscule probabilities or even the mere possibility of life ever starting or evolving on such a planetary system. But suppose life does form and evolve to higher forms, what would it be like to live on such a planet? We could rephrase the question: what sort of life we could expect to live if we were transported to such a planet. Of course such an interstellar travel being not really possible, it will have to be only in our imagination (a fantasy - Ooh La La!).

## What gives rise to a tidal locking?

The gravitational force of Moon raises a tidal bulge in the oceans on Earth, in general there being two tidal bulges on opposite sides of Earth. The bulges are there because of the differential gravitational force of Moon. The oceans at points closer to Moon experience a larger gravitational force than Earth's centre, and hence feel a larger attraction and the water is pulled away from the Earth toward the Moon, thereby producing a bulge on the near side. On the other hand the bulge on the far side arises because the water there is pulled toward Moon less than Earth's centre, therefore water there moving relatively away from Earth's centre. Therefore, as Earth rotates, a given point on the surface will experience two tides in a day.

Because of friction there is a delay in Earth's response, causing the tidal bulge to lead the Earth-Moon axis by a small angle. The Moon exerts a torque on the tidal bulge that retards Earth's rotation, thereby increasing the length of the day. This slowing down mechanism has been working for billion of years, since oceans first formed on the Earth. There is geological and paleontological evidence that the Earth rotated faster and that Moon was closer to Earth in remote past. Tidal rhythmites are alternating layers of sand and silt laid down offshore from estuaries having great tidal flows. Daily, monthly and seasonal cycles can be found in the deposits. The geological records are consistent with this that 620 million years ago the day was 21.9±0.4 hours and there were 400±7 solar days/year. Extrapolated backwards, the day was even shorter, with a year having perhaps more than a thousand days, during Earth's first billion years. Of course it is not the actual duration of the year that has changed, it is only the number of days that decreases, with the duration of each day becoming longer with the passage of time.

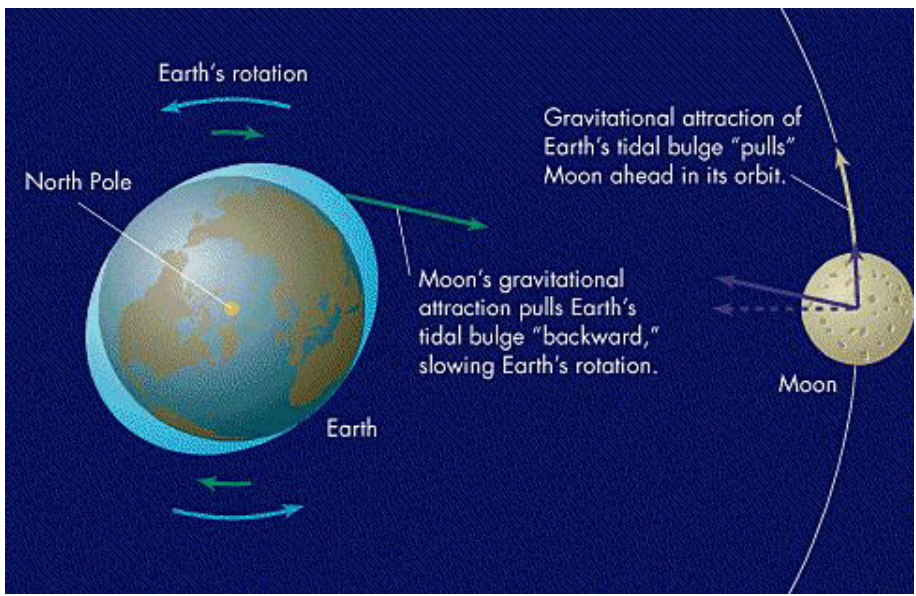

*Fig. 1: The slowing down of Earth's rotation due to tides caused by Moon's gravitational pull and the accompanying acceleration of Moon's orbital motion due the pull of Earth's tidal bulge.*

We have leap second once or twice a year (on 30th June or 31st December). The day lengthens by a second every 40,000 years. It may not seem much, but when linearly extrapolated it becomes substantial. The slowing rotation of the Earth results in a longer day as well as a longer month. The torque that Earth's tidal bulge exerts on the Moon leads to an acceleration of the Moon's orbital motion, causing the Moon to recede from Earth at an average rate of about 2 cm/year. The size of the orbit changes in such a way so as to conserve angular momentum for the system. If the Earth and Moon still continue to exist, the Moon's distance would have finally increased to about 1.35 times its current value by the time the length of the day equals the length of the month and the lunar tidal friction mechanism will cease. Earth will be then tidally locked to Moon. That has been projected to happen once the day and month both equal about 47 (current) days, billions of years in the future. Tides due to Sun will be still there, though the effect of the Sun on Earth tides is less than half that of the Moon. Eventually the rotation period of Earth might become equal to its revolution period around the Sun. The day-night cycle would having ceased, with one side of Earth always facing the Sun and the opposite side of Earth in a perpetual dark, and Earth thus getting tidally locked to Sun. Perhaps the time required for such a thing to happen may be too long and much before that Earth and Moon might get eventuality swallowed by Sun when it turns a red giant. But one can still imagine the possibility of such a tidal-locking to occur, if not here on Earth, may be elsewhere on another planet around some other star.

## Are there any tidally-locked planets somewhere?

Within the Solar system, apart from Moon there are many other satellites tidally locked with their primaries. Pluto and Charon are both tidally locked to each other. Close binary stars throughout the universe are expected to be tidally locked with each other. An unusual example is Tau Boötis, a star tidally locked by a planet. The time taken for a tidal locking actually depends on things like orbital distance, masses of both bodies, and the malleability of the orbiting objects. Generally, closer objects are more likely to experience tidal locking. For planets orbiting M-type stars, which are slightly smaller than our Sun, the region where planets become tidally locked overlaps with the so-called habitable zone, where water can remain as a liquid on a planet's surface. For such stars, we would expect that a lot of the planets in the habitable zone would be tidally locked.

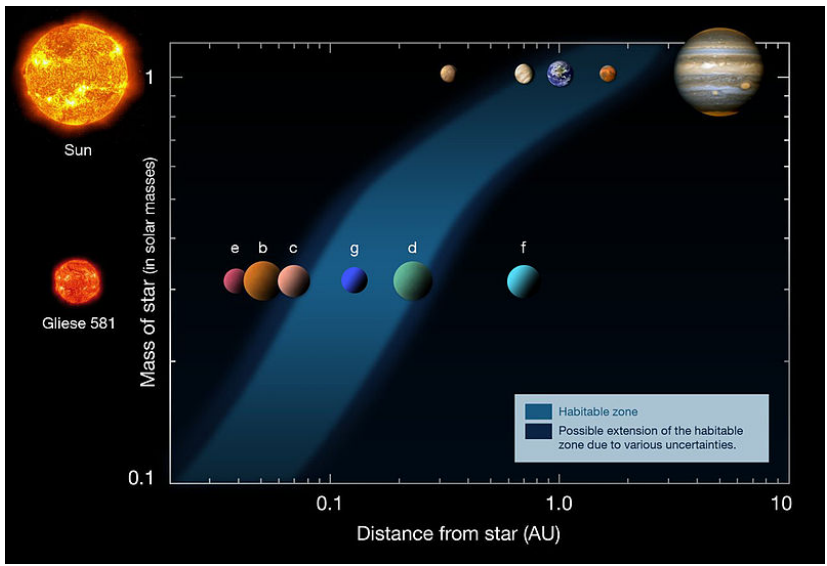

*Fig. 2: Planetary habitable zones of the Solar system and the Gliese 581 system compared.*

We now know that there are planetary systems around other stars, and already about a thousand or so such exoplanets have been found. There is evidence that some of them are tidally locked to their companion star. Zarmina is an extrasolar planet around Gliese 581, an M3V red dwarf star approximately 20.5 light-years away from Earth in the constellation of Libra. The planet lies near the habitable zone of its parent star, and the presence of liquid water is considered a strong possibility and is believed to be the most Earth-like planet, and the best exoplanet candidate with the potential for harbouring life found to date. There might be a finite possibility of life on such a planet. However, Zarmina is tidally locked to its star. The immediate disadvantage of a tidally-locked planet is obvious. One side of the planet cooks while the other freezes, playing havoc with the atmospheric system of the planet.

## Atmospheric system of a tidally-locked planet

On a tidally-locked planet, a single region is consistently close to the star. Known as the sub-stellar point, this region continuously receives more direct sunlight, and thus more heat. Water on one side may be in a vapour form while on the other side it may be frozen into ice. Strong constant heating of a planet on one side can change or even control how much weathering occurs on the planet, which can lead to significant and even unstable climate changes. These dramatic climate effects could make a planet that otherwise has the potential for life to instead be uninhabitable. An unstable climate on a tidally-locked planet could create a runaway greenhouse effect that could result in an atmosphere like that of Venus.

And it's not just the extreme temperatures. Permanently devoid of the heat of the star, the atmosphere on the dark side would first turn into a denser gas, then condense into a liquid, and then perhaps further condense into solid form. Meanwhile, air that is constantly exposed to light — or that is heated by a ground that is constantly exposed to light — will heat up and expand. Although it is doubtful whether the atmosphere on the dark side of the planet would get to solid form, it would certainly keep condensing and leaving a vacuum to suck in the expanding hot air from the other side. This might make for circulation of atmosphere that would make the planet liveable, but it will also lead to hellish storms, as the atmosphere from the light and dark side of the planet essentially switched sides continually. The presence of an atmosphere can help distribute the heat across the planet, equalizing the temperatures. But tidal locking could result in wide climate variations, a result that could threaten the evolution of life on the surface of these planets.

The far side of the planet would be frigid, since it would never see the star. Its only source of warmth would be winds from the warmer half of the planet. If there was no heat transport from the hot side to the cold side then we could expect the temperatures to be below $-100°$ C on the cold side, and more than 100° C on the hot side. However, if the planet has an atmosphere then it will transport heat from the hot side to the cold side, and this might make the temperature difference much more moderate.

Perhaps some of the water could be found in a liquid state near the boundary between the hot and cold regions and one could expect some kind of water cycle, with something like glaciers being continually melted by the warm air blowing in from the hot side, with the melted water flowing in gigantic rivers to the hot side, where it evaporates and cycles back around to fall as snow on the cold side.

## Could life be sustained on a tidally-locked planet?

Consider what life would be like for those living on a planet tidally locked to its star where while one side is permanently bright and boiling hot, the other side is permanently dark and freezing cold. The lit side of the planet will be stripped of its oceans and made to face burning star and scrubbing hot winds all the time, while the dark side will be covered with frozen oceans and biting cold winds howling all the time. Three would never be any relief from the extreme temperatures so it may be difficult to sustain life on such an inhospitable world. There would be zones of different climates, in concentric rings, depending upon how far away from the sub-stellar point. In the centre there would probably be scorching hot deserts but farther away from the sub-stellar point, as the star would get lower in the sky, there would be gradually cooler climates. Different places on the planet would be confined to either day or night, or even dusk at regions in between. However at a given place any change from day to night or vice versa will not occur.

Life, if it manages to struggle along on such a planet, will be very hard or perhaps be underground. More likely a circular belt between the two sides - a sort of "twilight zone" - could be the place for life to evolve and flourish. In this dusk band around the planet, where star will be permanently hanging very low near the horizon or perhaps the stellar disc partially peeking above the horizon, with an ever-colourful red, yellow sky due to scattered light, the temperatures would be more moderate, right in between the hot and cold sides. However the heat on one side would cause the air to rise, creating a low pressure system, while the cold on the other side would cause the air to sink, creating a high pressure system. This would cause the planet to experience a constant and violent circulation of air, or, essentially a planet-wide hurricane. The constant air circulation would actually circulate the temperatures extensively and extremes in temperature would mitigate. Water cycles with huge rivers crossing from cold to hot side might make living there possible.

One important issue is the concept of time. With no day-night cycles, concept of time will be difficult to come. On Earth right from birth, we notice that many phenomena in nature are repetitive. This is due to our most basic natural clock, viz. the rotation of the Earth, causing the rising and setting of the Sun, giving rise to alternative periods of light and darkness. All human and animal life has evolved accordingly, keeping awake during the day-light but sleeping through the dark night. Even plants follow a daily rhythm. Of course some crafty beings have turned nocturnal to take advantage of the darkness, e.g., the beasts of prey, blood–sucking mosquitoes, thieves and burglars, and of course astronomers.

At least there might be no astronomers on a tidally-locked planet, as starry sky may not be known (except for some rumours about it by adventurer fellows daring to venture deeper into darker side of the planet). Secrets of the Universe − planets, stars, Milky-way, other galaxies − all these might remain very difficult, if not outright impossible, to unravel. Just imagine, it took us humans thousands of year to figure out that a few "wandering stars" are heavenly bodies (planets) in just our neighbourhood and all this happened in spite of the fact that a starry sky is daily visible most places for about half of the time (night). How will the inhabitants of a tidally-locked planet ever know about it when their everlasting "hot summer afternoons" never turn into cool evenings and dark nights?

## A "day" in the life of a denizen of a tidally-locked planet

On Earth, early humans noticed that over a certain period of time, seasons changed, following an almost fixed pattern. Near the tropics - for instance, over most of India – a burning hot season gives way to torrential rains, in turn followed by months of bitter cold, then a couple of months of a pleasant weather springing up before the heat sets in again; the time period of this repetitive cycle defines a year. One would expect that on a tidally-locked Earth there would be no seasons. The only change in the amount of sunlight would come from the slight variation in distance from the Sun due to Earth's orbit being slightly elliptical. Certainly that would be the case if the axis of Earth were not inclined to the ecliptic.

So one could ask, will there be sort of seasons if the axis of a tidally-locked planet were tilted to its orbital plane like in case of Earth. Though there would be no day-night cycle on such a world, yet if the axis of the planet were inclined to its orbital plane around the star, there would be cycles of variation of temperature accompanying star's height over the horizon. This cycle would be the most natural clock for the denizens of this tidally-locked planet and thereby would define **day** on such a planet, regulating the daily life-cycle of the denizens of the planet. Then a "morning" will begin with a cold Winter, followed by a pleasant Spring and then a hot Summer afternoon, with the Fall falling in the evening, followed again by a Winter morning starting the next (year-long) day. Depending upon the latitude of the place there might occur a completely dark night in between or there might be no night ever.

Of course depending upon the width of the inhabitable belt, as the **day** will pass the temperature will vary, perhaps much more than what we are used to between various seasons on Earth. To escape the rising heat or the impending cold, the population would be forced to migrate towards darker or brighter sides, depending upon the time of the **day**, perhaps for hundreds or thousands of kilometres, depending upon the temperature variations and the size of the planet and width of the dusk band. We on Earth have day-night cycle with working or sleeping hours. With no day-night cycle, denizens of the tidally-locked planet would perhaps not evolve with sleep as a necessary ingredient of their daily life. Even on Earth there are perhaps many species, e.g. ants, which never sleep and work ceaselessly. May be denizens of the tidally-locked planet will also work incessantly their whole lives, migrating from one place to another without pause or stop. *Rest assured there might be no rest for the whole day on such a plant.* With no day-night cycle, no sleeping and waking hours, no office or school going daily-routines followed by leisurely weekends, with perhaps hardly any diversions and adding to it the drudgery of restless migrations from one place to another, the life might sound dull to us denizens of the present-day Earth, but to the denizens of the tidally-locked planet it might not sound so boring as they would have evolved within a very different system with a very different "life-style".

**Further Reading:**

Edwin S. Kite et al., *CLIMATE INSTABILITY ON TIDALLY LOCKED EXOPLANETS*, ApJ, 743, 41 (2011).

Sara Seager : *Is there life out there? – The search for habitable exoplanets, seagerexoplanets.mit.edu/ProfSeagerEbook.pdf*

Singal, T. & Singal, A. K., *Is interstellar space travel possible?* Planex News Lett. 3, issue 1, 22 (2013)

http://news.discovery.com/earth/what-would-happen-if-our-planet-became-tidally-locked-130202.htm
http://csep10.phys.utk.edu/astr161/lect/moon/tidal.html
http://csep10.phys.utk.edu/astr161/lect/time/tides.html
http://www.digipro.com/Trials/moon.html
http://physics.stackexchange.com/questions/29182/what-would-happen-if-the-earth-was-tidally-locked-with-the-sun